\documentclass[12pt]{amsart}
\usepackage{amscd}
\usepackage{amsfonts}
\usepackage{latexsym}

\begin{document}

\title[Finite Temperature Effects on Non-Commutative Manifolds]
{Effective Finite Temperature Partition Function for Fields on
Non-Commutative Flat Manifolds}
\author{A.A. Bytsenko}
\address{Departamento de Fisica, Universidade Estadual de Londrina,
Caixa Postal 6001, Londrina-Parana, Brazil\,\, {\em E-mail address:}
{\rm abyts@uel.br}}
\author{E. Elizalde}
\address{Instituto de Ciencias del Espacio (CSIC) \& Institut
d'Estudis Espacials de Catalunya (IEEC/CSIC), Edifici Nexus, Gran
Capit\`a 2-4, 08034 Barcelona, and Department ECM i IFAE, Facultat
de F\`isica,
Universitat de Barcelona, Diagonal 647, 08028 Barcelona, Spain\,\,
{\em E-mail address:} {\rm elizalde@ieec.fcr.es,
http://www.ieec.fcr.es/recerca/cme/eli.html}}
\author{S. Zerbini}
\address{Department of Physics, University of Trento,
Gruppo Collegato INFN di Trento, Sezione di Padova, Italy\;\;\;
\;\;\;\;\;\;\;\;\;\;\;\;\;\;\;\;\;\;\;\;\;\;\;\;\;\;\;\;\;\;\;\;
\;\;\;\;\;\;\;\;\;
{\em E-mail address:} {\rm zerbini@science.unitn.it}}

\date{March 8, 2001}

\thanks{AAB has been partially supported by a FAPESP grant (Brazil),
and by the Russian Foundation for Basic Research, grant No. 01-02-17157.
EE has been supported by DGDI (Spain),
project BFM2000-0810, by CIRIT (Generalitat de Catalunya),
contract 1999SGR-00257, and by the program INFN (Italy)--DGICYT (Spain).}

\begin{abstract}

The first quantum correction to the finite temperature partition function
for a self-interacting  massless
scalar field on a $D-$dimensional flat manifold with $p$ non-commutative
extra dimensions is evaluated by means of dimensional regularization,
suplemented with zeta-function techniques. It is found that the zeta
function associated with the effective one-loop operator may be nonregular
at the origin.
The important issue of the determination of the regularized vacuum energy,
namely the first quantum correction to the energy
in such case is discussed.

\end{abstract}

\maketitle

\bigskip


\bigskip

\section{Introduction}

Quantum field theories on non-commutative spaces generalize the familiar
structure of conventional field theories. A motivation for considering these
models is their appearance in $M-$theory and in the theory of strings
\cite{connes,douglas,seiberg,cheung,chu,schomerus,ardalan},
and also the fact that they are perfectly consistent formulations by
themselves. It has been  shown, for instance, that
non-commutative gauge theories describe the low energy excitations of open
strings on $D-$branes in a background Neveu-Schwarz two-form field $B$
\cite{connes,douglas,seiberg}.

Recently, the non-commutative perturbative dynamics on $D-$
dimensional manifolds have been investigated \cite{seiberg1} and
the Kaluza-Klein spectrum for interacting scalars and vector
fields has been calculated in \cite{gomis0}; it is consistent with
formulas found in \cite{seiberg1,kiem,fischler0}, where a
connection between calculations performed in string theory and
some others, done in field theory on a non-commutative torus, has
been made. The one-loop Casimir energy of scalar and vector fields
on a non-commutative space like ${\Bbb R}^{1,d}\otimes
T^2_{\theta}$ (where ${\Bbb R}^{1,d}$ is a flat
$(d+1)-$dimensional Minkowski space and $T^2_{\theta}$ a
two-dimensional non-commutative torus) has been calculated in
\cite{nam}. The higher dimensional non-commutative torus case has
been investigated in \cite{huang,bytsenko3}; termal effects in
perturbative non-commutative gauge theories have been actively
studied in Refs.
\cite{cai,arcioni,fischler,arcioni1,gomis,landsteiner}.

In this paper we will calculate - using  dimensional regularization
implemented with zeta-function techniques - the partition
function at inverse temperature $\beta$ for
massless scalar fields  defined on $D-$dimensional manifolds with compact and
non compact non-commutative dimensions. This can be considered as {\it the first}
quantum correction to the total partition function. Since we will  compute
a functional determinat of a one-loop operator, our computation is,
as well known,  equivalent to summing an infinite number of one-loop
diagrams \cite{seiberg1}.  The relevance of this contribution will be
discussed later.

The action of the massless interacting scalar fields we are
considering has the form $$ S_{({\rm
scalar})}=\int\left(\frac{1}{2}(\partial\,\phi)^2+\frac{\lambda}
{r!}\phi\star\phi\star...\star\phi\right)d^Dx \mbox{.}
\eqno{(1.1)} $$ As is well known, in ultrastatic spacetimes finite
temperature effects can be easily described with the help of the
imaginary time formalism - in which the fields are assumed to be
periodic with period $\beta$. For this reason, we will consider
manifolds with topology  $M=S^1\otimes{\Bbb R}^{d-1}\otimes
X_{\theta}^p$, where $p=D-d$ and ${\rm dim}\,M =D $.  In the non
compact case $X_\theta^p= {\Bbb R}^p_\theta$, while in the compact
case $X_\theta^p= T^p_\theta$ is a non-commutative torus. If we
denote by $x^\mu$ the non-commutative coordinates, we assume that
$$ [x^\mu,x^\nu]=i\theta^{\mu\nu} \mbox{,} \eqno{(1.2)} $$ with
$\mu,\nu=1,...p$, and $\theta^{\mu\nu}=\theta \sigma^{\mu\nu}$ is
a real, non-singular, antisymmetric matrix with entries $\pm1$,
being $\theta$  the  non-commutative parameter. $X_{\theta}^p$ may
be regarded as a non-commutative associative algebra, with elements
given by ordinary continuous functions on $X_{\theta}^p$ whose
product is given by the Moyal bracket or $(\star)-$product of
functions: $$ (F\star G )(x)=\exp\left(\frac{i}{2}\theta^{\mu\nu}
\frac{\partial}{\partial \alpha^\mu}\frac{\partial}{\partial
\beta^\nu} \right)F(x+\alpha)G(x+\beta)|_{\alpha=\beta=0} \mbox{.}
\eqno{(1.3)} $$ Since we have to deal with a renormalizable scalar
field theory,  the number $r$ of scalar fields in the
$(\star)-$product  should  be chosen as an integer, given by
$r=2D/(D-2)$. Clearly, the only possible
 choices are $D=3$, $r=6$,\,\,\,\,$D=4$, $r=4$, and $D=6$,
$r=3$.

We are interested in the one-loop approximation for a theory with
the action (1.1). It has been  shown that in the massless case,
the operator ${\mathcal L}$, related to the one-loop contribution,
has the formal aspect \cite{seiberg1}: $$ {\mathcal
L}=k^2+\frac{\Lambda}{k\circ k}+M^2 \mbox{,} \eqno{(1.4)} $$ where
$k^2$ is the $D-$dimensional Laplacian on ${\Bbb R}^D$ and $k\circ
k=\theta^2 k_\mu k_\mu$. The effective mass term $M^2$ depends on
$\lambda$ and  corresponds to the planar diagram contributions,
while the non local $\lambda $ term represents the non-planar
diagram contribution.

We are also assuming that the time coordinate is commutative
(unitarity of the field theory) and we denote by $\tau=-it$ the
compactified imaginary time with period $\beta$. Thus, the finite
temperature one-loop operator is $$ {\mathcal L}=L_{d}+L_p+\Lambda
(\theta^2 L_p)^{-1} \mbox{,} \eqno{(1.5)} $$ where
$L_{d}=-\partial^2_\tau+L_{d-1}=-\partial^2_\tau+k^2_j+M^2$ and $
L_p=k_\mu^2$ are the Laplacian-like operators on the spaces
$S^1\otimes {\Bbb R}^{d-1}$ and $X^p_\theta$, respectively.

We observe that in the {\it massive} case, the one-loop effective operator is
much more complicated. It reads
$$
{\mathcal L}=k^2+m^2+\frac{2\Lambda m}{(k\circ k)^{3/2}}
K_1(2m (k\circ k)^{-1/2})+M^2
\mbox{.}
\eqno{(1.6)}
$$
This appears as an untractable form. Since $K_1(z) \rightarrow 1/z$ as
$z \rightarrow  0$, in the massless limit, one obtains Eq. (1.4).
If we take the commutative limit, namely $\theta \rightarrow 0$ in Eq. (1.6),
one gets ${\mathcal L}=k^2+m^2+M^2$, which is the correct one-loop
operator in the commutative case.
In Ref. \cite{seiberg1}, an approximate expression valid for small $m$
has been considered and a mixing between UV and IR divergences has been
noticed. It should be noted that the massless limit and the limit
 $\theta \rightarrow 0$ do not commute, namely it is not possible to
consider the   $\theta \rightarrow 0$ in Eq. (1.4).
In this paper, for the sake of
simplicity, we shall consider the massless case only, bearing in mind
possible limitations of this choice.

\section{Regularization of the partition function}

We shall make use of dimensional regularization in the proper time formalism,
implemented by zeta-function techniques \cite{elizalde,elizalde2,bytsenko}.
The logarithm of the one-loop effective partition function is regularized
by  means of ($\mu$ is a renormalization parameter)
$$
{\rm log \,} Z_\beta(\varepsilon)=-\frac{1}{2}
\left({\rm log}\,{\rm det}\,{\mathcal L}/\mu^2\right)_{\varepsilon}=
\frac{1}{2}\mu^{2\varepsilon}\int_0^\infty dt t^{\varepsilon-1}
\mbox{Tr} e^{-t{\mathcal L}}
\mbox{,}
\eqno{(2.1)}
$$
where $\varepsilon$ is a small regularization parameter.
Since we are making use of the Gaussian approximation, what we are going to
compute is the first quantum correction to the partition function.

The object of interest is the heat kernel trace $\omega(t)$,
defined for $t>0$ by $\omega(t|{\mathcal L})={\rm Tr\, }e^{-t{\mathcal L}}$.
The zeta function $\zeta(s|{\mathcal L})$ and $\omega(t|{\mathcal L})$
are related by the Mellin transform:
$$
\zeta(s|{\mathcal L})=
\frac{1}{\Gamma(s)}\int_0^{\infty}\omega(t|{\mathcal L})t^{s-1}dt,\,\,\,\,\,
\mbox{for}\,\,\,\, \Re s>D/2
\mbox{.}
\eqno{(2.2)}
$$
Then, one has
$$
{\rm log \,} Z_\beta(\varepsilon)=\frac{1}{2}
\mu^{2\varepsilon} \Gamma(\varepsilon)\zeta({\varepsilon}|{\mathcal L})
\mbox{.}
\eqno{(2.3)}
$$
If the zeta function of the operator ${\mathcal L}$ is regular at
the origin, the one-loop divergences and the finite part of the
logarithm of the partition function can be expressed, respectively,
in terms of  the zeta function and its derivative evaluated  at the origin.
In this case the vacuum energy (Casimir energy) can be defined
resorting to the usual thermodynamical relation
$$
<E>\,\,\,=\,\,-\lim_{\beta \rightarrow \infty}\partial_\beta
{\rm log} Z_\beta
\mbox{,}
\eqno{(2.4)}
$$
where the renormalized partition function is simply
$$
{\rm log \,} Z_\beta=\frac{1}{2}\left(\frac{d}{ds}\zeta(s|
{\mathcal L})|_{s=0}
+{\rm log} \mu^2 \zeta(0|{\mathcal L})\right)
\mbox{.}
\eqno{(2.5)}
$$
For example, in the case $S^1 \otimes Y$, one gets (see Ref.
\cite{bytsenko})
$$
<E>\,\,\,=\,\,\frac{1}{2}{\rm PP}\,\zeta(-\frac{1}{2}|{\mathcal L}_Y)+
(1+{\rm log} 2\mu)\mbox{Res}\,\zeta(-\frac{1}{2}|{\mathcal L}_Y)
\mbox{,}
\eqno{(2.6)}
$$
where  ${\rm PP}$ denotes the principal part of the zeta function, given by
$$
{\rm PP}\,\zeta(-\frac{1}{2}|{\mathcal L}_Y)=\lim_{s \rightarrow 0}
\left( \zeta(s-\frac{1}{2}|{\mathcal L}_Y)+
\frac{A_{D/2}({\mathcal L}_Y)}{2\sqrt \pi s} \right)
\mbox{,}
\eqno{(2.7)}
$$
with ${\mathcal L}_Y$ being
the spatial operator acting on a manifold $Y$, namely ${\mathcal L}=-
\partial^2_\tau+{\mathcal L}_Y$, and $A_{D/2}$  the corresponding
heat-kernel coefficient. This prescription has been introduced in
\cite{blau} and rederived,
by using zeta-function regularization, in \cite{cognola92,kirsten1}.
In the following we will investigate the analytic continuation of
$\zeta(s|{\mathcal L})$.

Making use of heat-kernel techniques, one can show that the zeta
function of the one-loop operator may, in general, be nonregular
at the origin. Indeed, we have $$ \mbox{Tr\, }e^{-t {\mathcal L}}=
\mbox{Tr\, }e^{-t L_{d}}\mbox{Tr\, }e^{-t {\mathcal L}_X} \mbox{,}
\eqno{(2.8)} $$ where we have introduced the pseudo-differential
operator ${\mathcal L}_X=L_p+\Lambda(\theta^2 L_p)^{-1}$. The
first short $t$ asymptotics is standard and reads (for flat
manifolds) $$ \omega(t|L_{d})\simeq \frac{{\rm Vol}({\Bbb
R}^d)}{(4\pi t)^{d/2}}, \,\,\,\,\,\mbox {as} \,\, t\rightarrow
0^{+} \mbox{.} \eqno{(2.9)} $$

The second one involves the operator ${\mathcal L}_X$ and contains
logarithmic terms (see, for example, \cite{gilkey} and references
quoted there). For the cases we are interested in, we have $$
\omega(t|{\mathcal L}_X) \simeq \sum_{r=0}^\infty A_r t^{r-p/2}+
\sum_{k=0}^\infty B_k t^{2k+p/2} {\rm log \,} t \mbox{.}
\eqno{(2.10)} $$ Thus, the contribution  containing logarithmic
terms is $$ \omega(t|{\mathcal L})\simeq \frac{{\rm Vol}({\Bbb
R}^d)}{(4\pi)^{d/2}} \sum_{k=0}^\infty B_k t^{2k+(p-d)/2} {\rm
log} t+\mbox{non log terms} \mbox{.} \eqno{(2.11)} $$ Since a pole
of the zeta function at the origin corresponds to having the pure
log term ${\rm log}\,t$, it turns out that a pole will appear if
and only  if $d=p+4k\,,\,\,k\in {\Bbb Z}_+$ and  since $p$ is
even, $D$ also has to be even. The appearance of a pole at the
origin of the zeta function is quite unsual. It is in common with
quantum fields defined on higher dimensional cones
\cite{kirsten,cognola} and on 4-dimensional spacetimes with a non
compact (but of finite volume) spatial hyperbolic section
\cite{bytsenko4}.

One can calculate the effective partition function evaluating a regularized
functional
determinant of   $\zeta(s|{\mathcal L})$, Eq. (2.1), which can be rewritten as
$$
{\rm log} Z_\beta(\varepsilon)=\frac{1}{2}
\mu^{2\varepsilon}\Gamma(\varepsilon)\zeta({\varepsilon}|{\mathcal L})
\mbox{.}
\eqno{(2.12)}
$$
Recall we are going to compute a functional determinant, and thus we are
effectively adding up an infinite number of one-loop diagrams, planar and
non-planar ones. Our calculation should be viewed as an attempt at an
implementation of the background field method to the non-commutative case.

In order to go on, we need the evaluation of the analytic
continuation of the zeta function $\zeta(s|\mathcal L)$. Starting
from the Mellin transform and making use of the Poisson-Jacobi
resummation formula, a standard calculation leads to $$
\zeta(s|\mathcal L)=\frac{\beta}{(4\pi)^{\frac{1}{2}} \Gamma(s)}
\left(\Gamma(s-\frac{1}{2})\zeta(s-\frac{1}{2}|{\mathcal L}_Y)
\right. $$ $$ \left. + 2\sum_{n=1}^\infty \int_0^\infty dt
t^{s-\frac{3}{2}} e^{-\frac{n^2 \beta^2}{4t}} \omega(t|{\mathcal
L}_Y)\right) \mbox{.} \eqno{(2.13)} $$ Here ${\mathcal
L}_Y=L_{d-1}+{\mathcal L}_X$. In order to evaluate the high
temperature expansion the second term on the r.h.s. of Eq. (2.13)
can be conveniently rewritten as a Mellin-Barnes integral (see
\cite{bytsenko}). As a result, we have $$ \zeta(s|\mathcal
L)=\frac{\beta}{(4\pi)^{\frac{1}{2}}
\Gamma(s)}\left(\Gamma(s-\frac{1}{2})
\zeta(s-\frac{1}{2}|{\mathcal L}_Y) \right. $$ $$ \left. +
\frac{1}{2\pi i} \int_{\Re z=\frac{D}{2}}dz
\left(\frac{\beta}{2}\right)^{-z} \Gamma(\frac{z}{2})
\zeta_R(z)\Gamma(s+\frac{z-1}{2}) \zeta(s+\frac{z-1}{2}|{\mathcal
L}_Y) \right)   \mbox{,} \eqno{(2.14)} $$ where $\zeta_R(z)$ is
the Riemann zeta function. In the equations   above, the first
part represents  the vacuum contribution, while the second one is
the statistical sum contribution. It is also clear that one needs
to know the meromorphic structure of the zeta function
$\zeta(s|{\mathcal L}_Y)$, which can be obtained from
$\zeta(s|{\mathcal L}_X)$. Making an expansion in terms of the
effective mass $M^2$, a direct computation gives $$
\zeta(z|{\mathcal L}_Y)= \frac{{\rm Vol}({\Bbb
R}^{d-1})}{(4\pi)^{\frac{d-1}{2}}} \sum_{k=0}^\infty (-M^2)^k
\frac{\Gamma(z+k-\frac{d-1}{2})}{k!\Gamma(z)}
\zeta(z+k-\frac{d-1}{2}|{\mathcal L}_X) \mbox{.} \eqno{(2.15)} $$

\section{The spectral zeta function and the regularized vacuum energy}

First of all, let us consider the simple case of a non-compact,
non-commutative manifold $X_\theta^p={\Bbb R}_\theta^p$ (i.e., the
non-commutative Euclidean space). The heat kernel trace reads $$
\omega(t|{\mathcal L}_X)=\frac{2 {\rm Vol}({\Bbb
R}^p)}{(4\pi)^{p/2} \Gamma(p/2)}a^{p/2}K_{p/2}(2at) \mbox{,}
\eqno{(3.1)} $$ where $K_\nu(z)$ is the Mac Donald function and
$a^2=\Lambda \theta^{-2}$. It is easy to show that the short$-t$
asymptotics of $\omega(t|{\mathcal L}_X)$ is of the kind given by
Eq. (2.10). The Mellin transform leads to $$ \zeta(z|{\mathcal
L}_X)=\frac{{\rm Vol}({\Bbb R}^{p})}{2(4\pi)^{p/2}}
\frac{\Gamma(\frac{2z+p}{4})\Gamma(\frac{2z-p}{4})}{\Gamma(p/2)\Gamma(z)}
a^{p/2-z} \mbox{.} \eqno{(3.2)} $$ Eq. (3.2) gives the analytic
continuation of the zeta function.

In the case of a compact non-commutative  manifold  (i.e., the
non-commutative tori) $X_\theta^p=T_\theta^p$, for large $\Re\,s$,
the spectral zeta function  associated with the operator
${\mathcal L}_X$ reads \cite{gomis,nam,bytsenko3}: $$
\zeta(s|{\mathcal L}_X)= \sum_{{\bf n}\in {\Bbb Z}^p/{\{{\bf
0}\}}} \varphi({\bf n})^{-s} \left[ 1+ \Lambda \theta^{-2}R^{4}
\varphi({\bf n})^{-2}\right]^{-s} \mbox{,} \eqno{(3.3)} $$ where
$R$ is the compactification radius. The analytic continuation can
be achieved by using binomial expansion
(\cite{eejmp2,elizalde,elizalde2,bytsenko}). As a result, for
$C=a^2R^{4}<1$ one has $$ \zeta(s|{\mathcal L}_X) = R^{2s}
\sum_{\ell=0}^{\infty} \frac{(-C)^{\ell}\Gamma\left(s+\ell
\right)} {\ell!\Gamma\left( s\right)} Z_p\left| \begin{array}{ll}
{\bf 0}\\ {\bf 0}\\
\end{array} \right|\left(2s+4\ell \right)
\mbox{,}
\eqno{(3.4)}
$$
where  $Z_p\left|_{\bf h}^{\bf g}\right|(s,\varphi)$ is the $p$-dimensional
Epstein zeta function associated with the quadratic form
$\varphi [a({\bf n}+{\bf g})]=\sum_ja_j(n_j+{\rm g}_j)^2.$ For $\Re\,s>p$
\,\, ,
$Z_p\left|_{\bf h}^{\bf g}\right|(s,\varphi)$ is given by the formula
$$
Z_p\left| \begin{array}{ll}
{\rm g}_1\,...\,{\rm g}_p \\
h_1\,...\,h_p\\
\end{array} \right|(s,\varphi)=\sum_{{\bf n}\in {\bf Z}^p}{}'
\left(\varphi[a({\bf n}+{\bf g})]\right)^{-s/2}
\exp\left[2\pi i({\bf n},{\bf h})\right]
\mbox{,}
\eqno{(3.5)}
$$
where ${\rm g}_j$ and  ${\rm h}_j$ are some real numbers \cite{bateman},
and the prime means omitting the term with $(n_1, n_2, ..., n_p)=
({\rm g}_1, {\rm g}_2, ..., {\rm g}_p)$ if all of the ${\rm g}_j$ are
integers.   The functional equation for
$Z_p\left|_{\bf h}^{\bf g}\right|(s,\varphi)$ reads
$$
Z_p\left| \begin{array}{ll}
{\bf g}\\
{\bf h}\\
\end{array} \right|(s,\varphi)=
({\rm det}\,a)^{-1/2}\pi^{\frac{1}{2}(2s-p)}\frac{\Gamma(\frac{p-s}{2})}
{\Gamma(\frac{s}{2})}\exp[-2\pi i({\bf g},{\bf h})]
$$
$$
\!\!\!\!\!\!\!\!\!\!\!\!\!\!\!
\times
Z_p\left| \begin{array}{ll}
{\bf\,\,\,\, h}\\
-{\bf g}\\
\end{array} \right|(p-s,\varphi^*)
\mbox{,}
\eqno{(3.6)}
$$
where $\varphi^*[a({\bf n}+ {\bf g})]=\sum_j a_j^{-1}(n_j+{\rm g}_j)^2$.
Formulas (3.4) and (3.6) give the analytic continuation of the zeta function.

Alternatively, a powerful expression for the analytic
continuation, that extends (to arbitrary number of dimensions and
to more general Epstein-like zeta functions) the Chowla-Selberg
formula for the homogeneous Epstein zeta function in
two-dimensions, has been given recently \cite{eli1,eli2,eli3}. It
provides an expression valid in the whole of the complex plane,
exhibiting the poles explicitly, and under the form of a
power-fast convergent series. For the compact case, Eq. (3.3), the
formula reads: $$ \zeta(s|{\mathcal L}_Y)= \frac{2^{s-d+2}{\rm
Vol}({\Bbb R}^{d-1})}{(2\pi)^{(d-1)/2} \Gamma(s)}
\sum_{\ell=0}^\infty \sum_{j=0}^{p-1}\frac{(-4\Lambda
\theta^{-2})^\ell \Gamma (s+\ell-\frac{d-1}{2})}{\ell! \, \Gamma
(s+2\ell-\frac{d-1}{2}) \left(\det{{\bf A}_{j}}\right)^{1/2}} $$
$$ \times \left[ \pi^{j/2} a_{p-j}^{-s-2\ell +(d+j-1)/2} \Gamma
(s+2\ell-\frac{d+j-1}{2}) \zeta_R(2s+4 \ell-d-j+1)  \right. $$ $$
\left. + 4 \pi^{s+2\ell-(d-1)/2} a_{p-j}^{-s/2-\ell-(d+j-1)/4}
\sum_{n=1}^\infty \sum_{{\bf m}_j \in {\Bbb
Z}^j}\!\!'n^{(d+j-1)/2-s-2\ell}\right. $$ $$ \left. \times
\left({\bf m}_j^t {\bf A}_j^{-1}{\bf
m}_j\right)^{s/2+\ell-(d+j-1)/4} K_{(d+j-1)/2-s-2\ell}\left(2\pi n
\sqrt{ a_{p-j}{\bf m}_j^t {\bf A}_j^{-1}{\bf m}_j}\right) \right]
\mbox{,} \eqno{(3.7)} $$ where $a_j=R_j^{-2}$, $R_j$ being the
compactification radius corresponding to the $j$th coordinate
(above, and in what follows, only the particular case $R_j =R,
\forall j$, is considered). Note that the term $j=0$ in the second
sum reduces to: $a_p^{s-2\ell} \Gamma (s+2\ell)
\zeta_R(2s+2\ell)$, which coincides with the result of the
1-dimensional case ($p=1$), as it should. Here ${\bf A}_j$ is the
submatrix of ${\bf A} =\mbox{diag\,} (a_1,\ldots , a_p)$ made up
of the last $j$ rows and columns: ${\bf A}_j=\mbox{diag\, }
(a_{p-j+1},\ldots , a_p)$.

Now we have in our hands all the ingredients necessary to study the limit
$\varepsilon$ goes
to $0$ in Eq. (2.3), making use of Eqs. (3.4) and (3.6).
The vacuum contribution to the one-loop effective action has been
investigated  recently in \cite{bytsenko3}.
The presence of poles at the origin in the zeta function of Eq. (3.4)
(see also Ref.
\cite{bytsenko3}) imposes a re-analysis of the renormalization. This
can proceed as follows.

For small $\varepsilon$ one has
$$
{\rm log} Z_\beta(\varepsilon)=\frac{1}{2}
\mu^{2\varepsilon}\Gamma(\varepsilon)\zeta({\varepsilon}|{\mathcal L})
=\frac{W_0}{\varepsilon^2}+\frac{W_1}{\varepsilon}+W_2+{\mathcal O}
(\varepsilon)
\mbox{.}
\eqno{(3.8)}
$$
It is convenient to introduce the ``regular'' zeta function
$$
\xi(s|{\mathcal L})=s\zeta(s|{\mathcal L})
\mbox{.}
\eqno{(3.9)}
$$
Multiplying both sides of Eq. (3.8) by $\varepsilon^2$, we obtain the
following expressions for the coefficients $W_j$:
$$
W_0=\frac{1}{2}\xi(0|{\mathcal L})=\frac{1}{2}{\rm Res}
\,\left(\zeta(\varepsilon|{\mathcal L})|_{\varepsilon =0}\right)\,
\mbox{,}
\eqno{(3.10)}
$$
$$
W_1=\frac{1}{2}{\rm lim}_{\varepsilon\rightarrow 0}\frac{d}{d\varepsilon}
\left(\mu^{2\varepsilon} \Gamma(1+\varepsilon)
\xi(\varepsilon|{\mathcal L})\right)
$$
$$
=\frac{1}{2}\left(\frac{d}{ds}\xi(s|{\mathcal L})|_{s=0}+({\rm log} \mu^2-\gamma)
\xi(0|{\mathcal L})  \right)
\mbox{,}
\eqno{(3.11)}
$$
$$
W_2=\frac{1}{4}{\rm lim}_{\varepsilon\rightarrow 0}\frac{d^2}{d^2
\varepsilon}  \left(\mu^{2\varepsilon} \Gamma(1+\varepsilon)
\xi(\varepsilon|{\mathcal L})\right)
$$
$$
=\frac{1}{4}\left[\frac{d^2}{ds^2}\xi(s|{\mathcal L})|_{s=0}+
2({\rm log} \mu^2-\gamma)\frac{d}{ds}\xi(s|{\mathcal L})|_{s=0}\right]
$$
$$
+ \frac{\xi(0|{\mathcal L})}{4}\left[({\rm log} \mu^2-\gamma)^2
+\frac{d}{ds}\Psi(1+s)|_{s=0} \right]
\mbox{,}
\eqno{(3.12)}
$$
where  $\Psi(s)$ is the logarithmic derivative of the Gamma
function and $\gamma=-\Psi(1)$ is the Euler-Mascheroni constant.
The coefficients $W_0$ and $W_1$ are the counterterms, while
$W_2$ is the renormalized partition function. If the zeta function
$\zeta(0|{\mathcal L})$ is regular at the origin, the quantity defined by
Eq. (3.12) reduces to the renormalized partition function, because
$$
\xi(0|{\mathcal L})=0\,,\,\,\,\,\,\,
\frac{d}{ds}\xi(s|{\mathcal L})|_{s=0}=\zeta(0|{\mathcal L)},
$$
$$
\frac{d^2}{ds^2}\xi(s|{\mathcal L})|_{s=0}=
\frac{d}{ds}\zeta(s|{\mathcal L})|_{s=0}
\mbox{.}
\eqno{(3.13)}
$$

\subsection{The regularized vacuum energy}

In order to evaluate the regularized vacuum energy, one needs the meromorphic
structure of the zeta functions
$\zeta(z|{\mathcal L})$ and $\zeta(z|{\mathcal L}_X)$. The latter  can
 be obtained by making use of the short heat
kernel expansion (2.9). A direct calculation gives $$
\Gamma(z)\zeta(z|{\mathcal L}_X)=\sum_{r=0}^\infty
\frac{A_r}{z+r-p/2}-\sum_{k=0}^\infty
\frac{B_k}{(z+2k+p/2)^2}+J(z) \mbox{,} \eqno{(3.14)} $$ where
$J(z)$ is the analytic part. To be noted is the presence of second
order poles associated with the logarithmic terms. Thus, $$
\!\!\!\!\!\!\!\!\!\!\!\!\!\!\!\!\!\!\!\!\!\!\!\! \Gamma(z)
\zeta(z|{\mathcal L}_Y)= \sum_{k=0}^\infty \frac{(-M^2)^k}{k!}
\left( \sum_{r=0}^\infty \frac{A_r({\mathcal
L}_Y)}{z+k+r-\frac{D-1}{2}}\right. $$ $$ \left. -\sum_{k=0}^\infty
\frac{B_k({\mathcal L}_Y)}{(z+2k+\frac{p-d+1}{2})^2}
+J(z|{\mathcal L}_Y)\right) \mbox{,} \eqno{(3.15)} $$ where we
have introduced the Seeley-De Witt coefficients $$ A_r({\mathcal
L}_Y)= \frac{{\rm Vol}({\Bbb R}^{d-1})}{(4\pi)^{\frac{d-1}{2}}}A_r
,\,\,\,\,\,\,\,\,\,\, B_r({\mathcal L}_Y)= \frac{{\rm Vol}({\Bbb
R}^{d-1})}{(4\pi)^{\frac{d-1}{2}}}B_r \mbox{.} \eqno{(3.16)} $$

If $D$ is even and if $\zeta(s|{\mathcal L})$ has a pole  at the origin,
Eqs. (2.4), (2.14), (3.12) and (3.15) lead to
$$
\!\!\!\!\!\!\!\!\!\!\!\!\!\!\!\!
\!\!\!\!\!\!\!\!\!\!\!\!
<E>\,\,\,=\,\,\frac{1}{2}{\rm PP}\,\zeta(-\frac{1}{2}|{\mathcal L}_Y)+
(1+{\rm log}2\mu-\frac{\gamma}{2})\mbox{Res}\,
\zeta(-\frac{1}{2}|{\mathcal L}_Y)
$$
$$
\:\:\:\:\:\:\:\:\:\:\:\:\:\:\:\:
+\frac{1}{4}B_{(D-2p)/4}({\mathcal L}_Y)
\left[{\rm log}^2 \mu^2-\left(\frac{\pi^2}{6}
+(2-2{\rm log}2-\gamma)^2\right)\right]
\mbox{,}
\eqno{(3.17)}
$$
where now
$$
\!\!\!\!\!\!\!\!\!\!\!\!\!\!\!\!\!\!\!\!\!\!\!\!\!\!\!\!
\!\!\!\!\!\!\!\!\!\!\!\!\!\!\!\!\!\!\!\!\!\!\!\!\!\!\!\!
\!\!\!\!\!\!\!\!\!\!\!\!\!\!\!\!\!\!\!\!\!\!\!\!\!\!\!\!
{\rm PP}\zeta(-\frac{1}{2}|{\mathcal L}_Y)
= \lim_{s \rightarrow 0}\left\{
\zeta(s-\frac{1}{2}|{\mathcal L}_Y)\right.
$$
$$
\left.
-\frac{1}{2\sqrt \pi}\left[
\frac{B_{(D-2p)/4}({\mathcal L}_Y)}{s^2}
-\frac{1}{s}
[A_{D/2}({\mathcal L}_Y)
+\Psi(-\frac{1}{2})B_{(D-2p)/4}({\mathcal L}_Y)] \right]\right\}
\mbox{.}
\eqno{(3.18)}
$$
The coefficients $B_{(D-2p)/4}({\mathcal L}_Y)$ are related to  the
residue of $\zeta(s|{\mathcal L})$ at the origin.
This is the {\it new} prescription for the evaluation of the regularized
vacuum energy. When there is no pole at the origin for $\zeta(s|{\mathcal
L})$,  the prescription reduces to the standard one given by (2.7).

\subsection{Non-Commutative Euclidean space}

In our situation, i.e. in the case of massless scalar field defined on a
$D$ manifold
with noncommuting coordinates, the relevant choices are $D=3,\, 4,\, 6$.

For the sake of simplicity, let us consider only the  non compact case.
In the case $D=3$,\,
$d=2,\,\,\,\, \zeta(s|{\mathcal L})$ is regular
at the origin and furthermore $\zeta(0|\mathcal L)=0$. Thus the vacuum
energy  is given simply by
$$
\!\!\!\!\!\!\!\!\!\!\!\!\!\!\!\!\!\!\!\!\!\!
\!\!\!\!\!\!\!\!\!\!\!\!\!\!\!\!\!\!\!\!\!\!\!\!\!\!\!\!\!\!\!\!
\!\!\!\!\!\!\!\!\!\!\!\!\!\!\!\!\!\!\!\!\!\!\!\!\!\!\!\!\!\!\!\!
\!\!\!\!\!\!\!\!\!\!\!\!\!\!\!\!\!\!\!\!\!\!\!\!\!\!\!\!\!\!\!\!
<E>\,\,=\,\frac{1}{2}\,\zeta(-\frac{1}{2}|{\mathcal L}_Y)
$$
$$
\:\:\:\:\:\:\:\:\:\:\:\:\:\:\:\:\:\:\:\:
=
\frac{{\rm Vol}({\Bbb R}^2)a^{3/2}}{32 \pi^{3/2}}
\sum_{k=0}^\infty \frac{ (-\tilde{M}^2)^k}{k!}
\Gamma\left(\frac{2k-1}{4}\right)\Gamma\left(\frac{2k-3}{4}\right)
\mbox{,} \eqno{(3.19)} $$ while for $D=6$, $d=2$ and $p=4$, from
Eq. (2.6)  we have $$ <E>\,\,=-\,\frac{{\rm Vol}({\Bbb
R}^5)a^3}{256\pi^3} \left[ \sum_{k=0,\,\,k \neq 1,3}^\infty
\frac{(-\tilde{M}^2)^k}{k!}
\Gamma\left(\frac{k-1}{2}\right)\Gamma\left(\frac{k-3}{3}\right)\right.
$$ $$ \left. -\tilde{M}^2 \frac{\tilde{M}^4-6}{12} \left(1+ 2\pi
\left({\rm log}\,\,\frac{\mu^2}{2a}+ \frac{1}{2\sqrt
\pi}(\Psi(1)-\Psi(1/2)\right)\right) \right] \mbox{.}
\eqno{(3.20)} $$ In Eqs. (3.19) and (3.20) $\tilde{M}^2\equiv
M^2/a$ and this factor is dimensionless and independent on the
coupling constant.

The situation changes for $ D=6$\, ($d=4$,\,\, $p=2$) and  $D=4$\,
($p=d=2$). In these cases the zeta function has a pole at the
origin and one has to make use of Eq. (3.17). Let us first
consider the case $D=6$,\, $d=4$,\, $p=2$.  We have $$
<E>\,\,=-\,\frac{{\rm Vol}({\Bbb R}^5)a^3}{256\pi^3} \left[
\sum_{k=0,\,\,k \neq 1,3}^\infty \frac{(-\tilde{M}^2)^k}{k!}
\Gamma\left(\frac{k-1}{2}\right)\Gamma\left(\frac{k-3}{2}\right)\right.
$$ $$ \left.+ \tilde{M}^2 \left(\frac{1}{2} \Psi'(1)+ \left({\rm
log}\,\,\frac{\mu^2}{a}-\gamma \right)^2+2+
(4-\frac{\tilde{M}^4}{3}) \left({\rm
log}\,\,\frac{\mu^2}{a}-\gamma \right)\right) \right] \mbox{.}
\eqno{(3.21)} $$ In the other case, $D=4$,\,\,$p=d=2$,\, Eq.
(3.17) gives $$ <E>\,=-\,\frac{{\rm Vol}({\Bbb R}^3)a^2}{64\pi^2}
\left[ \sum_{k=1,\,\,k\neq 2}^\infty \frac{ (-\tilde{M}^2)^k}{k!}
\Gamma\left(\frac{k}{2}\right)\Gamma\left(\frac{k-2}{2}\right)
-2-\frac{4\pi^2}{3}\right. $$ $$ \left. +8\gamma^2
-8\left(\frac{1}{2}-2\gamma + {\rm
log}\,\,\frac{\mu^2}{a}\right)^2 + \tilde{M}^4 ({\rm
log}\,\,\frac{\mu^2}{2a}+\frac{1}{2}\Psi(1)-\frac{1}{2}\Psi(1/2))\right]
\mbox{.} \eqno{(3.22)} $$ Our evaluation of the functional
determinant corresponding to the thermal fluctuation operator has
also led to a non-analytic dependence on the coupling constant
$\lambda$, corresponding to the resummation of all the one-loop
diagrams. As is well known, higher-loop diagrams could, in
principle, give a contribution of the same order. In any case, our
computation is the necessary first step towards the final result.

In the compact case, we only obsserse that we have qualitatively the same
results, but they are much more involved.
Here, the sign of vacuum energy is relevant for issues concerning
the stabilization
mechanism of the compactification radius. However, it is to be noticed that
 its value  depends on the
renormalization parameter $\mu$ and on the number of non commuting
coordinates.

\subsection{High temperature expansion}

To obtain the high temperature expansion, $\beta \rightarrow 0$, one may use
Eqs. (2.14) and (3.15). We consider only the case $D=4$
since for the other cases, $D$ odd or $D=6$, the high temperature
expansion is
standard, in its first leading terms.

For suitable $s$, we shift the vertical contour in the second term of Eq.
(2.14) to the left.
There are several simple poles and double poles in $z$. The simple pole
at $z=0$ gives a contribution which cancels with the first term on the r.h.s.
of Eq. (2.14). Then there are simple poles at $z=1$ and $z=4-2r-2s$. The
double poles occur at $z=-4k-2s$. As a result, the thermal zeta function
reads
$$
\zeta(s|\mathcal L) = \zeta(s|{\mathcal L}_Y)+
\frac{\beta}{(4\pi)^{\frac{1}{2}}  \Gamma(s)}
\left[ \left(\frac{\beta}{2}\right)^{-4}A_0({\mathcal L}_Y)\zeta_R(4)
\right.
$$
$$
\left.
+ \left(\frac{\beta}{2}\right)^{-2}
A_1({\mathcal L}_Y)\zeta_R(2)
+A_2({\mathcal L}_Y) \zeta_R(-1)+{\mathcal O}(s)\right]
$$
$$
- \frac{\beta^{2s+1} B_0({\mathcal L}_Y)}{(2\pi)^{2s+1} \Gamma(s)}
\left[ -{\rm log}\left(\frac{\beta}{2\pi}\right)\Gamma(s+1/2)\zeta_R(1+2s)
\right.
$$
$$
\left.
-\frac{1}{2}\frac{d}{ds}\Gamma(s+1/2)\zeta_R(1+2s)
-\Gamma(s+1/2)\frac{d}{ds}\zeta_R(1+2s) \right]
+\frac{{\mathcal O}(\beta^5)}{\Gamma(s)}
\mbox{.}
\eqno{(3.25)}
$$
It should be noted that the high temperature expansion is still singular
at $s=0$ due to the  $(d/ds)\zeta_R(1+2s)$ term, and, strictly speaking, one
should make use of Eq. (3.12) in order to compute the first quantum correction
to the  partition  function. However, this term
originates  a subleading contribution, of order zero in $\beta$,
the leading contributions being similar to those in the non-commutative case.

\section{Conclusions}

In this paper we have calculated the first quantum correction to the
partition function at finite temperature
for massless scalar fields on flat manifolds with non compact and with
 compact non-commutative dimensions.  Dimensional regularization
implemented with zeta-function techniques has been used to perform the
calculations, which are rendered absolutely rigorous through the use
of such machinery. This is not a trivial issue. In fact, the
difficulties posed by the appearance of the non-commutative contributions
in the spectral functions (which depend on it non analytically) should not be
underestimated.  Our calculation should be viewed as an attempt at an
implementation of the background field method to the non-commutative case.

On the positive side, this has as a consequence that the quantum theory at
finite temperature obtained by `addition' of the non-commutative dimensions,
departs appreciably from the commutative case. This is true concerning the
mathematics of the problem: new poles in the zeta function, new asymptotic
expansions, as well as the physics
of the model. Here we have investigated the changes suffered by the
vacuum energy density in the cases considered.
After obtaining the general formula for the family of spacetimes at issue,
a new prescription for the regularized vacuum energy has been obtained.

This new prescription has been exemplified with the
explicit expressions corresponding to dimensions $D=3,\,4,\,6$,
where changes
in the coefficients of the corresponding expressions of the regularized
vacuum energy density are notorious.

One should also notice that, in the case of the presence of the pole at the
origin for the zeta function, the dependence on the
renormalization
parameter in the vacuum energy (or in the effective action) is quite
complicated, and, as a consequence, the one-loop renormalization group
equations will be different from the standard ones.
Furthermore, making use of the Eq. (A.10), contained in the Appendix of
Ref. \cite{eli4}, it is possible to show that the use of a different
regularization, for example the cut-off regularization, leads to the same
conclusions.

 We have also discussed
the high temperature expansion for the partition function and have shown
that new subleading terms, associated with the non-commutative part of the
operator spectrum, are present.
In subsequent work we aim at exploring the subject further, by trying to
approach an experimental situation where one could actually check the
non-commutative results vs the ordinary, commutative ones.

\end{document}